\documentclass{ws-ijmpe}
\usepackage[super,compress]{cite}

\newcommand{\bk}{\mbox{\boldmath $k$}}
\newcommand{\br}{\mbox{\boldmath $r$}}

\newcommand{\bx}{\mbox{\boldmath $x$}}
\newcommand{\by}{\mbox{\boldmath $y$}}

\newcommand{\bsigma}{\mbox{\boldmath $\sigma$}}

\newcommand{\brho}{\mbox{\boldmath $\rho$}}

\newcommand{\bks}{\mbox{\scriptsize \boldmath $k$}}

\newcommand{\beqa}{\begin{eqnarray}}
\newcommand{\eeqa}{\end{eqnarray}}
\newcommand{\nn}{\nonumber}

\begin{document}

\markboth{S. Aoki, J. Balog, T. Doi, T. Inoue, P. Weisz}{Short Distance Repulsion Among Baryons}

\catchline{}{}{}{}{}

\title{Short Distance Repulsion Among Baryons}

\author{\footnotesize SINYA AOKI\footnote{Address after April 1, 2013: Yukawa Institute for Theoretical Physics, Kyoto University,  Kitashirakawa Oiwakecho, Sakyo-ku, Kyoto 606-8502, Japan}}

\address{Graduate School of Pure and Applied Sciences, University of Tsukuba, Ten-oh-dai 1-1-1\\
Tsukuba, Ibaraki 305-8571, Japan
\\
saoki@het.ph.tsukuba.ac.jp}

\author{JANOS BALOG}

\address{Institute for Particle and Nuclear Physics, Wigner Research Centre for Physics, H-1525 \\
Budapest 114, P.O.B. 49, Hungary
\\
balog.janos@wigner.mta.hu}

\author{TAKUMI DOI}

\address{Theoretical Research Division, Nishina Center, RIKEN \\
Wako 351-0198, Japan
\\
doi@ribf.riken.jp}

\author{TAKASHI INOUE}

\address{Nihon University, College of Bioresource Sciences \\
Kanagawa 252-0880, Japan
\\
inoue.takashi@nihon-u.ac.jp}

\author{PETER WEISZ}
\address{Max-Planck-Institut f\"ur Physik, F\"ohringer Ring 6, \\
D-80805 M\"unchen, Germany
\\
pew@mpp.mpg.de}

\maketitle

\begin{history}
\received{Day Month Year}
\revised{Day Month Year}
\end{history}

\begin{abstract}
We review recent investigations on the short distance behaviors of potentials among baryons, which are formulated through the Nambu-Bethe-Salpeter (NBS) wave function.
After explaining the method to define the potentials, we analyze the short distance behavior of the NBS wave functions and the corresponding potentials by combining the OPE (operator product expansion) and a renormalization group analysis in the perturbation theory of QCD. These analytic results are compared with numerical results obtained in lattice QCD simulations.
\end{abstract}


\ccode{PACS numbers: 12.38.Gc,  12.38.Bx, 13.75.-n,21.30.-x}


\section{Introduction}
A force between nucleons, the nuclear force, is the most fundamental quantity in nuclear physics.
The phenomenological nucleon-nucleon (NN) potentials~\cite{NN-review} exhibit long-to-medium distance attractions,
which have been explained by meson exchanges between nucleons~\cite{yukawa},
while they have short distance repulsion, the repulsive core\cite{jastrow}, essential for the stability of nuclei against collapse, whose origin has not yet been well understood.
Moreover, general short distance repulsions among baryons, not only NN but also three-nucleon (3N), baryon-baryon (BB) and three-baryon (3B)\cite{Nishizaki:2002ih}, 
may be required to explain a recently observed\cite{Demorest:2010bx} neutron star as heavy as twice a solar mass . Unfortunately it is difficult or almost impossible to determine these short distance repulsions experimentally, except for the NN case, while 
the theoretical determination  for these short distance repulsions requires difficult non-perturbative calculations in QCD.

Recently, a novel method  to define the NN potential in QCD has been proposed~\cite{Ishii:2006ec,Aoki:2008hh,Aoki:2009ji} (we call it  the HAL QCD scheme or method in this paper), and the method has been widely applied to various hadronic 
interactions~\cite{Ishii:2009zr,Murano:2011nz,Nemura:2008sp,Nemura:2012fm,Sasaki:2010bi,Ikeda:2011qm} 
including BB potentials~\cite{Inoue:2010hs,Inoue:2010es} and 3N forces~\cite{Doi:2011gq}.
Furthermore, in the HAL QCD scheme, the short distance behavior of two- and three baryon potentials can be investigated analytically via the operator product expansion combined with a renormalization group (RG) analysis of perturbative QCD\cite{Aoki:2010kx,Aoki:2011aa,Aoki:2009pi,Aoki:2010uz,Aoki:2012xa}.

In this paper, we review both analytical investigations and numerical results for short distance behaviors of these potentials in the HAL QCD method . In Sec.~\ref{sec:method} we briefly explain the HAL QCD method to define the potential in QCD, taking the NN case as an example.
The OPE (operator product expansion) formalism to analyze the short distance behaviors of potentials in the HAL QCD scheme is summarized in Sec.~\ref{sec:OPE}.
Results for potentials in lattice QCD simulations are summarized and compared with the OPE analyses in
Sec.~\ref{sec:results}, and a summary is given in Sec.~\ref{sec:conclusion}. 

\section{HAL QCD method}
\label{sec:method}
In this section, we explain the HAL QCD method to define the potential in QCD for the NN case. 
The basic quantity is the equal-time Nambu-Bethe-Salpeter(NBS) wave function, defined by
\beqa
\varphi^{\bks,s_1s_2}_{\alpha\beta}(\bx) &=& \langle 0 \vert T\left\{N_\alpha(\bx_0, 0) N_\beta(\bx_0 + \bx, 0) \right\}\vert NN,\bk, s_1s_2 \rangle_{\rm in}, 
\label{NBSwavefn}
\eeqa
where $\langle 0 \vert $ is the QCD vacuum (bra)-state, 
$ \vert NN, \bk, s_1 s_2\rangle_{\rm in} $ is the two-nucleon asymptotic in-state with helicity $s_1$, $s_2$
and  relative momentum $\bk$ in the center of mass frame, whose total energy is given by 
$W_{\bks}=2\sqrt{\bk^2+m_N^2}$ with the nucleon mass $m_N$, and 
$T$ represents the time-ordered product.
The local interpolating operator for the nucleon 
is taken as $ N_\alpha(x) =\varepsilon_{abc} (u^a(x)^T C\gamma_5 d^b(x) ) 
q^c_\alpha(x)$ (with $x=(\bx,t)$), 
where $a,b,c$ are the color indices, $\alpha$ is the spinor index, $C=\gamma_2\gamma_4$ is the charge conjugation matrix,   
and $q(x) =(u(x),d(x))$. 

As long as the total energy $W_{\bks}$ lies below the pion production threshold({\it i.e.} $W_{\bks} < W_{\rm th}=2 m_N + m_\pi$
with pion mass $m_\pi$),
the NBS wave function at  $r=|\bx| \rightarrow \infty $ satisfies the Helmholtz equation:
\begin{equation}
\left[ {k^2} +\nabla^2 \right] \varphi^{\bks,c}_{\Gamma}(\bx )\simeq 0, \qquad k =\vert \bk \vert ,
\end{equation}
where $c\equiv s_1 s_2$ and $\Gamma\equiv\alpha \beta$.
Moreover,  the radial part of the NBS wave function for
given orbital angular momentum $L$ and total spin $S$ for asymptotically large $r$
becomes\cite{Ishizuka2009a,Aoki:2009ji}
\begin{eqnarray}
 \varphi^{\bks}(r;LS) \propto 
\frac{\sin( k r - L\pi/2 + \delta_{LS}(k))}{k r}  e^{i\delta_{LS}(k)} ,
\label{eq:asympt} 
\end{eqnarray}
 where $\delta_{LS}(k)$ is the NN phase shift obtained from the S-matrix in QCD below the inelastic threshold. 
 
From the above property of the NBS wave function, it is possible to define a non-local potential  through the equation
\begin{eqnarray}
\left[ E_k- H_0\right] \varphi^{\bks,c}_{\Gamma}(\bx ) &=&
\sum_{\Gamma_1}\int U_{\Gamma; \Gamma_1}(\bx, \by)  \varphi^{\bks,c}_{\Gamma_1}(\by )d^3y, \  
 \left( E_k=\frac{k^2}{2\mu},\  H_0=-\frac{\nabla^2}{2\mu} \right) 
\label{eq:schroedinger}
\end{eqnarray}
with the reduced mass  $\mu=m_N/2$, where $U(\bx,\by)$ is expected to be short-ranged because of absence of massless particle exchanges between two nucleons.
The potential $U(\bx,\by)$ is finite and does not depend on the particular renormalization scheme,   
since the NBS wave function $\varphi^{\bks,c}_{\alpha\beta}$ is multiplicatively renormalized. 
While Lorentz covariance is lost by using the equal-time NBS wave function and 
Eq. (\ref{eq:schroedinger}) is written as a Schr\"{o}dinger type equation, 
a non-relativistic "approximation" is NOT introduced to define $U(\bx,\by)$.

The non-local but energy($\bk$)-independent potential $U(\bx,\by)$  can be formally given
by\cite{Aoki:2008hh,Aoki:2009ji}, 
\begin{eqnarray}
U_{\Gamma_1,\Gamma_2}(\bx,\by) &=&  \sum_{\bks_1,c_1\bks_2,c_2}^{\vert\bks_1\vert, \vert\bks_2\vert < k_{\rm th}} \left[E_{k_1} - H_0\right]\varphi^{\bks_1,c_1}_{\Gamma_1}(\bx) {\cal N}^{-1}_{\bks_1,c_1; \bks_2,c_2}\varphi^{\bks_2,c_2}_{\Gamma_2}(\by)^\dagger ,
\label{eq:U}
\end{eqnarray}
where $k_{\rm th} $ is the threshold momentum defined to satisfy $2\sqrt{k_{\rm th}^2+m_N^2} = W_{\rm th}$ so that the summation over
$\bk_1$, $\bk_2$ is restricted below the inelastic threshold, and
${\cal N}^{-1}$ is the inverse of ${\cal N}$ defined from the inner product of NBS wave functions as 
\begin{eqnarray}
{\cal N}^{\bks_1,c_1; \bks_2,c_2} &=&\left( \varphi^{\bks_1,c_1}, \varphi^{\bks_2,c_2} \right)
\equiv \sum_\Gamma \int d^3x\ \varphi^{\bks_1,c_1}_\Gamma(\bx)^\dagger\, \varphi^{\bks_2,c_2}_\Gamma (\bx)
\end{eqnarray}
for $\vert \bk_1\vert, \vert \bk_2\vert < k_{\rm th}$.  This $U(\bx,\by)$ is energy-independent by construction and satisfies eq.~(\ref{eq:schroedinger})
as
\begin{eqnarray}
\sum_{\Gamma_1}\int U_{\Gamma; \Gamma_1}(\bx, \by)  \varphi^{\bks,c}_{\Gamma_1}(\by )d^3y &=&
\sum_{\bks_1,c_1\bks_2,c_2}^{\vert\bks_1\vert, \vert\bks_2\vert < k_{\rm th}} \left[E_{k_1} - H_0\right]\varphi^{\bks_1,c_1}_{\Gamma}(\bx) {\cal N}^{-1}_{\bks_1,c_1; \bks_2,c_2}\cdot {\cal N}^{\bks_2,c_2; \bks,c} \nonumber \\
&=&  \left[E_{k} - H_0\right]\varphi^{\bks,c}_{\Gamma}(\bx) 
\end{eqnarray}
as long as $W_{\bks} < W_{\rm th}$.
Once the non-local potential $U(\bx,\by)$ which satisfies eq.~(\ref{eq:schroedinger}) is obtained,
eq.~(\ref{eq:asympt}) guarantees that  we can extract the phase shift 
$\delta_{LS}(k)$ at  $W_{\bks} < W_{\rm th}$
in QCD by solving the Schr\"odinger equation with this potential. 
Note however that the non-local potential which satisfies eq.~(\ref{eq:schroedinger}) at $W_{\bks} < W_{\rm th}$
is not unique, since the potential itself is not a physical observable.
We may add terms orthogonal to all NBS wave functions below the inelastic threshold, which only affect eq.~(\ref{eq:schroedinger}) at $W_{\bks} \ge W_{\rm th}$\cite{Aoki:2009ji}.

The construction of $U(\bx,\by)$ in eq.~(\ref{eq:U}) is formal and is inadequate for practical use, since
only a limited number of wave functions at low energies (ground state and possibly a few low-lying excited  states)
can be obtained  in lattice QCD simulations in a finite box. In practice, we therefore 
expand the non-local potential  in terms of the velocity (derivative) with 
 local coefficient functions\cite{TW67};
$
U(\bx,\by) = V(\bx,\nabla) \delta^3(\bx-\by), 
$
whose leading order terms reads
\begin{eqnarray}
V^{\rm LO}(\bx) = V_0(r) +  V_\sigma(r) \bsigma_1\cdot\bsigma_2 + V_T(r) S_{12},
\label{eq:LO}
\end{eqnarray}
where $\bsigma_i$ is the Pauli-matrix acting on the spin index of the $i$-th nucleon,  and
$
S_{12} = 3 (\bx\cdot \bsigma_1) (\bx\cdot \bsigma_2) /r^2 -\bsigma_1\cdot\bsigma_2
$
is the tensor operator. 
It has been found in numerical simulations\cite{Murano:2011nz} that
contributions from higher order terms are much smaller than those from $V^{\rm LO}$ at low energy.  This means that non-locality is rather small at low energy in our definition of the potential. 
 
For example for $L=0$ and $S=0$, we obtain
\begin{equation}
V_C(r, L=S=0)\equiv V_0(r) - 3V_\sigma(r) = 
\frac{\left(E_k -H_0\right)\varphi^{\bks, c}(r, L=S=0)}
{\varphi^{\bks,c}(r,L=S=0)} .
\end{equation}

As mentioned before, since the potential itself is not a physical observable, 
its short-distance behavior depends on its definition, in particular, the choice of nucleon operator
in the NBS wave function.
In this review, we  analyze the short distance behaviors of the potentials defined in the HAL QCD scheme exclusively, by comparing analytic results to numerical results from lattice QCD simulations.
The method in this review, however, can be easily applied to other definitions for the potential. 
 
\section{OPE and short distance behavior of the potential}
\label{sec:OPE}

\subsection{Notation and Method}

One of the outstanding properties of QCD is asymptotic freedom; 
renormalized running couplings $\bar{g}^2(r)$ which
characterize interactions at short distance scales $r$
tend logarithmically to zero as the scale decreases
$
\bar{g}^2(r)\sim -\frac{1}{2\beta_0\ln(\Lambda r)}\,,
$
where $\beta_0=\left[11-2N_f/3\right]/(16\pi^2)$ 
for gauge group SU(3), provided the number of
dynamical quarks is limited ($N_f\le 16$).
This has the important consequence that the 
behavior of a wide class of physical quantities
at short distances (alternatively high energies)   
can be computed perturbatively.
The NBS wave functions defined in eq.~(\ref{NBSwavefn}) 
belong to this class.
In the framework of perturbation theory of QCD 
it is convenient to regularize the 
ultra-violet
divergences using dimensional regularization (DR), and then 
bare parameters and operators must be renormalized. 

The operator product expansion (OPE) method is based on the observation that the product of the two renormalized nucleon operators appearing
in the definition of the NBS wave function
is for short distances expressed as a sum 
of local gauge invariant operators times coefficient functions: 
\begin{equation}
N_\alpha(0,0) N_\beta(\bx, 0)\mathop{\sim}_{\bx\to0}
\sum_A F^A_{\alpha\beta}(\bx)\mathcal{O}_A\,,
\end{equation}
which holds when sandwiched between asymptotic states
or in correlation functions with fundamental fields
at widely separated distances.
The coefficient functions $F^A_{\alpha\beta}(\bx)$
appearing above can be computed perturbatively using 
the RG equations which express independence of the
bare theory on the renormalization scale.
\footnote{For details we refer the reader to sect.~3
of ref.~\cite{Aoki:2010kx}.}
If the nucleon operators and the $\mathcal{O}_A$ 
are chosen so that they renormalize multiplicatively 
without mixing:
$
\mathcal{O}_A=Z_A\mathcal{O}_{A,\mathrm{bare}}\,,
$
then the corresponding coefficients behave as 
$
F^A_{\alpha\beta}(\bx)\sim \bar{g}^{2(\ell_A-\beta_A)}(|\bx|)\,,
$
where $\ell_A$ determines the loop order at which the 
corresponding operator first appears in perturbation theory (PT)
and $\beta_A$ is related to the difference of the 1-loop
anomalous dimension $\gamma_A$ of the operator $\mathcal{O}_A$ 
and that of two nucleons $\gamma_N=1/4\pi^2$: $\beta_A=(\gamma_A-2\gamma_N)/(2\beta_0)$.

The leading asymptotic short distance behavior of the NBS 
wave function is then dominated by the operator 
with the smallest value of $\ell_A-\beta_A$,
provided of course that its matrix element between the vacuum 
and the particular 2N state does not vanish 
(which is generically the case).
Usually the smallest value of $\ell_A-\beta_A$
is attained for operators $\mathcal{O}_A$
which appear at tree level ($\ell_A=0$) i.e. 
$c_A\ne0$ in the free field expansion:
\begin{eqnarray}
N_{\alpha;\mathrm{bare}}(0,0)N_{\beta;\mathrm{bare}}(\bx,0)
\sim_{g_0=0}&& 
:N_{\alpha;\mathrm{bare}}N_{\beta;\mathrm{bare}}:(0,0)+\mathrm{O}(\bx)
\nonumber \\
&=&\sum_B c_B\mathcal{O}_{B;\mathrm{bare}}(0,0)+\mathrm{O}(\bx)\,.
\end{eqnarray}

Given an arbitrary basis of gauge invariant $r$-quark operators
the anomalous dimensions are related to the eigenvalues
of the matrix $Z$ specifying the mixing under renormalization.
The renormalization constant matrix can be
determined by computing the correlation
function of the operators multiplied by $r$ quark fields
at widely separated distances. The 1-loop computation
just involves the diagrams with a gluon line connecting 
a pair of quark lines emanating from the operator vertex.
The divergent part, using DR 
in $D=4-2\epsilon$ dimensions is symbolically of the form:
\begin{equation}
\left[ q^{a,f}_\alpha(x) q^{b,g}_\beta (x)\right]^{\rm 1-loop, div}
= \frac{g^2}{96\pi^2}\frac{1}{\epsilon} 
\left[ ({\bf T}_0 + \lambda {\bf T}_1) \cdot 
q^a(x)\otimes q^b (x)\right]_{\alpha,\beta}^{f,g}
\label{eq:1-loopT}
\end{equation}
where ${\bf T}_0\,,{\bf T}_1$ are matrices in flavor and spinor space.
The term proportional to the covariant gauge parameter $\lambda$
cancels with the renormalization of the external quark lines.
Thus only ${\bf T}_0$ is relevant here, which in a chiral representation
of the gamma matrices, i.e. $\gamma_5=\mathrm{diag}(1,1,-1,-1)\,,$ 
is simply given by
\begin{eqnarray}
({\bf T}_0)^{f f_1,g g_1}_{\alpha\alpha_1,\beta\beta_1} &=&
\delta^{ff_1}\delta^{gg_1}
\left[ \delta_{\alpha\alpha_1}\delta_{\beta\beta_1}
    - 2\delta_{\beta\alpha_1}\delta_{\alpha\beta_1}\right]
+3\delta^{gf_1}\delta^{fg_1}
\left[ \delta_{\beta\alpha_1}\delta_{\alpha\beta_1}
-2\delta_{\alpha\alpha_1}\delta_{\beta\beta_1} \right]\nn \\
\label{eq:T0}
\end{eqnarray}
where either $\alpha_1,\beta_1 \in \{1,2\}$(right-handed) or
$\alpha_1,\beta_1 \in \{3,4\}$(left-handed).

Now any local gauge invariant 6--quark operator of (lowest) 
canonical dimension 9
can be written as a linear combination of operators of the form
\begin{equation}
B^{F_1,F_2}_{\Gamma_1,\Gamma_2}(x)\equiv
B^{F_1}_{\Gamma_1}(x) B^{F_2}_{\Gamma_2} (x)\,,\,\,\,
B^F_\Gamma (x) \equiv B^{fgh}_{\alpha\beta\gamma}(x) 
= \varepsilon^{abc} q^{a,f}_\alpha(x) q^{b,g}_\beta(x) q^{c,h}_\gamma (x)\,,
\,\,\,\,
\label{sixqops}
\end{equation}
where $\Gamma=\{\alpha,\beta,\gamma\}$ are sets of spinor labels and 
$F=\{f,g,h\}$ are sets of flavor labels.

Due to the structure of (\ref{eq:T0}), 
which reflects the chirality conservation of massless QCD,
the operators in eq.~(\ref{sixqops}) mix only with operators 
which preserve the set of flavor and Dirac indices in the chiral basis i.e.
$$
F_1\cup F_2 = F'_1\cup F'_2\,,\,\,\,
\Gamma_1\cup\Gamma_2 = \Gamma'_1\cup\Gamma'_2\,.
$$
We thus start by constructing sets of operators with fixed flavor 
and Dirac structure. 
Note however that such operators are not all linearly independent.
Relations between them of the form
\begin{equation}
3 B^{F_1,F_2}_{\Gamma_1,\Gamma_2} +
\sum_{i,j=1}^3 B^{(F_1F_2)[i,j]}_{(\Gamma_1,\Gamma_2)[i,j]} = 0\,,
\label{eq:constraint}
\end{equation}
follow from an identity satisfied by 
the totally antisymmetric $\varepsilon$ symbol and the 
Grassmannian nature of the quark fields.  
Here the $i$-th index of $abc$ and $j$-th index of $def$ are interchanged
in $(abc,def)[i,j]$. For example,
$(\Gamma_1,\Gamma_2)[1,1]=\alpha_2\beta_1\gamma_1,\alpha_1\beta_2\gamma_2$ or
$(\Gamma_1,\Gamma_2)[2,1]= \alpha_1\alpha_2\gamma_1,\beta_1\beta_2\gamma_2$.
Note that the interchange of indices occurs simultaneously for both.

For a given set of Dirac and flavor indices the initial set of operators
may be quite large. The determination of an independent basis
and the diagonalization of the renormalization constant matrix
$Z$ is then more conveniently done using algebraic
computer programs written in Mathematica or Maple.

Although the NBS wave function is always dominated by
the operator with the largest value of $\beta=\beta_C$,
for the potential this is only the case if $\beta_C\ne0$.
The potential is repulsive
at short distances if $\beta_C<0$ and attractive if $\beta_C>0$.
However if $\beta_C=0$ the leading term in the potential is
governed by the next largest non-zero anomalous dimension, 
and the sign of the potential requires non-perturbative
additional information. That there are operators
with $\beta_C=0$ is clear; 
e.g. operators of the form $B^{ffg}_{\alpha[\beta,\alpha]}
B^{ggf}_{\hat{\alpha}[\hat{\beta},\hat{\alpha]}}$ 
where $\alpha,\beta$  have the same chirality and
$\hat{\alpha},\hat{\beta}$ the opposite chirality,
since there is no contribution from diagrams where 
the gluon line joins quarks in the different baryon operators.

We have carried out the program above for the case
of 2-baryon wave functions involving 3 different
flavors of quarks and also for 3-baryon wave functions.
The main results from these analyses are 
presented in the next subsection.

\subsection{Results}

We have considered the following four problems: short-distance interaction 
between two nucleons \cite{Aoki:2010kx}, between two octet baryons 
\cite{Aoki:2010uz}, among three nucleons \cite{Aoki:2011aa} and finally
among three octet baryons \cite{Aoki:2012xa}. In all these cases we mainly
concentrated on the question of the existence of a repulsive (or attractive) core of the
interaction potential.

\subsubsection{2-body forces}

The technical part of the computation \cite{Aoki:2010kx} for these cases 
consists of three main steps: the enumeration of independent (taking into 
account the identities (\ref{eq:constraint})) local gauge invariant 6-quark
operators, calculation of the spectrum of 1-loop anomalous dimensions (by
diagonalizing the 1-loop mixing matrix) and finally checking whether
a given operator appears already at tree level in the OPE.
We found that all 1-loop anomalous dimension eigenvalues are, when multiplied
by $48\pi^2$, even integers. This suggests that there is a rationally related
natural basis in which the mixing matrix is automatically diagonal but we 
were unable to find a simple method to find this basis.

\medskip 

\noindent
{\it 2-nucleon potential (2 flavors) \cite{Aoki:2010kx}}

\noindent
The most important property of the spectrum in this case is that for
all operators
\begin{equation}
\gamma_A\leq2\gamma_N
\label{ineq1}
\end{equation}
hence $\beta_A$, the leading power of the running coupling is non-positive.
Almost all operators have negative $\beta_A$ indicating repulsive behavior,
but as explained above, there are also operators with the equality sign
in (\ref{ineq1}). In these cases whether the leading behavior corresponds to
attraction or repulsion cannot unfortunately be decided using perturbation
theory alone. In the short distance limit the potential behaves as
\begin{equation}
V(\bx)\sim R\,\frac{{\bar g}^{2(1-\beta^\prime)}(\vert\bx\vert)}{\bx^2}
\label{V2N}
\end{equation}
in these cases, where $\beta^\prime<0$ corresponds to the subleading operator
and $R$ is the ratio of the matrix elements of the leading
and subleading operators sandwiched between the vacuum and the given 2-nucleon
state. The latter can only be calculated non-perturbatively. Since we are
only interested in the {\it sign} of this quantity, we were able to use
a simple effective model to conclude \cite{Aoki:2010kx} that this sign is 
always positive leading to repulsion in all cases.

\medskip

\noindent
{\it 2-baryon potential (3 flavors) \cite{Aoki:2010uz}}

\noindent
The tensor product of two octets can be decomposed under SU$(3)$ as
\begin{equation}
8\otimes8=1_s\oplus8_s\oplus27_s\oplus8_a\oplus10_a\oplus\overline{10}_a,
\end{equation}
where the subscript $s(a)$ indicates symmetry (antisymmetry) under the
interchange of the two octet  representations. We considered local 6-quark
operators with flavor content $uuddss$ because such operators represent all
multiplets in the above decomposition. Again, most operators have $\gamma_A
\leq2\gamma_N$ but there are also attractive cases, operators with
$\gamma_A>2\gamma_N$ (see Table \ref{BBtab}). 
These all belong to the representations $1_s$, $8_s$ and
$8_a$, consistent with what we found for the 2-nucleon case, since 2-nucleon
states belong to the $\overline{10}_a$ and $27_s$ SU$(3)$ 
representations. Finally we found that all operators with $\gamma_A>2\gamma_N$
appear already at tree level in the OPE of two nucleon fields.
\begin{table}[tb]
\caption{List of channels with 1-loop anomalous
dimensions greater than $2\gamma_N$;
others can be obtained by symmetry transforms 
of the Dirac indices $1\leftrightarrow 2\,,
3\leftrightarrow 4$ or $(1,2)\leftrightarrow(3,4)$.}
\label{BBtab}
\begin{center}
\begin{tabular}{|c|c|c|}
\hline
Dirac structure & $48\pi^2\gamma_A$ & SU(3) representation \\
\hline
$112334$ & $42$ & $1$\\
$111222$ & $36$ & $1$\\
$112234$ & $36$ & $1$\\
$111234$ & $32$ & $1\oplus 8$\\
$112234$ & $32$ & $1\oplus 8$\\
\hline
\end{tabular}
\end{center}
\end{table}

\subsubsection{3-body forces}

As explained in the introduction, collective phenomena in nuclear physics
cannot be described using only 2-body potentials. In particular, the
existence of massive neutron stars suggests that also the 3-nucleon potential
has a repulsive core. To study 3-body forces with our method we introduce the
3-particle generalization of the NBS wave function (\ref{NBSwavefn})
\beqa
\label{NBSwavefn3}
\varphi^E_{\alpha\beta\gamma}(\br,\brho) 
&=& \langle 0 \vert T\left\{B_\alpha(\bx_1, 0) B_\beta(\bx_2, 0) 
B_\gamma(\bx_3,0)\right\}
\vert 3B,E \rangle_{\rm in}\,, 
\eeqa
where for transparency we suppressed momentum and helicity dependence. 
Here $B_\alpha$ is the baryon-octet generalization of the
nucleon doublet field appearing in (\ref{NBSwavefn}) and $3B$ is an
appropriate 3-baryon state with energy $E$. In the argument of the
3-body NBS wave function we use Jacobi coordinates $\br=\bx_1-\bx_2$
and $\brho=[\bx_3-(\bx_1+\bx_2)/2]/\sqrt{3}$.

In the local potential approximation we have
\begin{equation}
\left[-\frac{1}{m_N}\left(\nabla^2_r+\frac{1}{4}\nabla^2_\rho\right)+{\cal V}_{3B}\right]
\varphi^E_{\alpha\beta\gamma}=E\varphi^E_{\alpha\beta\gamma}.
\end{equation}
This defines the total 3-baryon potential ${\cal V}_{3B}$ and the true
3-baryon potential $V_{3B}$ is given by the potential excess
\begin{equation}
{\cal V}_{3B}(\br,\brho)=\sum_{i<j} V_{2B}(\bx_i-\bx_j)+V_{3B}(\br,\brho),
\end{equation}
where $V_{2B}$ is the 2-baryon potential discussed above.

In the UV limit, i.e. when $s=\sqrt{\br^2+\brho^2}\to0$ we can use the
OPE method to calculate the leading behavior of the wave function
(\ref{NBSwavefn3}) analogously to the 2-body case but here we have to 
consider local gauge invariant 9-quark operators. We have carried out the
anomalous dimension analysis both in the 3-nucleon case and the
most demanding 3-baryon case.

\medskip

\noindent
{\it 3-nucleon potential (2 flavors) \cite{Aoki:2011aa}}

\noindent
The most striking feature of the spectrum in this case is that for
{\it all} operators
\begin{equation}
\gamma_A<3\gamma_N.
\label{NNNres}
\end{equation}
Here again operators with the largest anomalous dimension are present in the
3-nucleon OPE at lowest order and we conclude that the total local potential
in this case behaves asymptotically as 
\begin{equation}
{\cal V}_{3N}(\br,\brho)\sim\frac{{\bar g}^2(s)}{s^2}
\end{equation}
(up to a positive constant). Since at short distances this  dominates
the 2-nucleon potential which behaves as given in (\ref{V2N}) it follows 
that the genuine 3-body potential $V_{3B}$ introduced above is responsible 
for the short-distance repulsion.

We note that showing the presence of a repulsive core in the 3-nucleon case
is the most important result of our perturbative considerations.
Universal repulsion follows unambiguously from (\ref{NNNres}) without having 
to rely on any model-dependent results.

We also note that one can observe from the calculated pattern of the 
spectrum of 1-loop anomalous dimensions that the Pauli exclusion
principle is at work here: increasing the number of quarks while keeping the
number of flavors fixed makes repulsion stronger while increasing the number
of flavors and fixing the number of quarks brings in some attractive
channels.

\medskip

\noindent
{\it 3-baryon potential (3 flavors) \cite{Aoki:2012xa}}

\noindent
In this case the SU$(3)$ decomposition is
\beqa
8\otimes 8\otimes  8 &=&
 64 \oplus ( 35 \oplus \overline{35})_2 \oplus  27_6 \oplus 
( 10\oplus \overline{10})_4 \oplus 8_8 \oplus 1_2 
\eeqa
and one can show that it is sufficient to work with the flavor content
$uuudddsss$ since all representations on the right hand side of the above 
formula occur in this flavor sector. 

The 3-baryon case is the most demanding technically. There are many Dirac index
configurations to be considered and for some of them the dimension of the
problem is quite large: with Dirac index structure 111223344 for example
there are originally 14130 operators and this number is reduced to 1518 after
imposing the 9-quark generalization of the identities (\ref{eq:constraint}).
After performing the diagonalization in all Dirac sectors we found that in the 
3-baryon case, although the vast majority of the eigenvalues are $<3\gamma_N$,
there are also some attractive channels. We enumerate these cases in 
Table \ref{BBBtab} (only operators appearing at tree level in the OPE
are listed).

Our calculations allow us to determine the leading UV behavior of the
total 3-baryon potential ${\cal V}_{3B}$. However, to draw any conclusion
concerning the UV limit of the genuine 3-body potential $V_{3B}$ is
difficult since comparing the results in Tables \ref{BBtab} and \ref{BBBtab}
we see that in some cases the 2-body forces are dominant, moreover there are 
also cases where there is cancellation between two different 2-body
contributions. 

\begin{table}[tb]
\caption{List of 3-baryon channels with anomalous dimensions greater than 
or equal to $3\gamma_N$. Only operators present at tree level are listed.}
\label{BBBtab}
\begin{center}
\begin{tabular}{|c|c|c|}
\hline
Dirac structure & $48\pi^2\gamma_A$ & SU(3) representation \\
\hline
$111223344$ & $48$ & $8$\\
            & $44$ & $1,8$\\
            & $42$ & $1,8$\\
            & $36$ & $8, 10, \overline{10}$ \\
$111333224$ & $44$ & $1,8$\\
$111222334$ & $48$ & $8$\\
            & $44$ & $1$\\
$111133442$ & $42$ & $1,8$\\
$111122334$ & $44$ & $1$\\
\hline
\end{tabular}
\end{center}
\end{table}

\section{Results from lattice QCD and comparisons with OPE predictions}
\label{sec:results}

In lattice QCD simulations with the spatial volume $L^3$, the NBS wave function can be extracted from the 4-point correlation function\cite{Ishii:2006ec,Aoki:2009ji}, which is 
given,  for example, in the case of the $NN$ system at $t > t_0$ by
\begin{eqnarray}
{\cal G}_{\alpha\beta}(\bx, t-t_0) &\equiv & \langle 0 \vert T\{N_\alpha(\br,t) N_\beta(\br+\bx,t)\}
\overline{\cal J}_{NN}(t_0)\vert 0\rangle \\
&=& \sum_{n,s_1,s_2} A(\bk_n,s_1,s_2)\  \varphi^{\bks_n,s_1,s_2}_{\alpha\beta}(\bx)e^{-W_{\bks_n} (t-t_0)} +\cdots \\
&\longrightarrow & A(\bk_0,s_1,s_2)  \varphi^{\bks_0,s_1,s_2}_{\alpha\beta}(\bx) e^{-W_{\bks_0} (t-t_0)} , \quad
t\gg t_0 , 
\end{eqnarray}
with the matrix element $A(\bk_n,s_1,s_2) ={}_{\rm in}\langle NN, \bk_n,s_1,s_2\vert  \overline{\cal J}_{NN}(0)\rangle$, where $ \overline{\cal J}_{NN}(t_0)$ is some source operator which can create states $\vert NN, \bk_n, s_1,s_2\rangle_{\rm in}$, and the ellipses represent inelastic contributions from intermediate states other than $NN$.  
In our numerical simulations, we have mainly used a wall source with the Coulomb gauge fixing only at $t=t_0$\cite{Aoki:2009ji}.   The large $t-t_0$ necessary   
to achieve ground state saturation in the last line causes some difficulties in numerical simulations.  
The signal-to-noise ratio for nucleon 4-pt functions behaves asymptotically for large $t$ as\cite{Parisi:1983ae,Lepage:1989hd} 
$
\displaystyle
\left(\frac{\cal S}{\cal N}\right) \sim e^{-2(m_N-3m_\pi/2)t},
$
where $m_N$ and $m_\pi$ are the nucleon and pion masses, respectively.
This problem becomes worse as we decrease  the pion mass toward its physical value. 
Furthermore,  as we increase the volume, the splitting between the ground state and the 1st excited state for the 2 nucleon system becomes smaller as
$
\displaystyle
\Delta E \simeq \frac{\bk_{\rm min.}^2}{m_N} = \frac{1}{m_N}\left(\frac{2\pi}{L}\right)^2.
$
If $L\simeq 5fm$ and $m_N\simeq 1$ GeV,  $\Delta E \simeq 62 {\rm MeV} \simeq 1/(3.2{\rm fm})$, which requires $ t \gg (\Delta E)^{-1}\simeq 3.2$ fm for ground state saturation.
The behavior of statistical noise in the above, however, makes the signals very poor at such large $t$ for the two nucleon system.

Recently, an improved extraction of the NBS wave function has been proposed to overcome the above difficulties\cite{HALQCD:2012aa}.
We first define the normalized 4-point correlation function as
\begin{eqnarray}
R(\bx,t) &\equiv& {\cal G} (\bx,t)/ (e^{-m_N t} )^2 
\simeq \sum_{n,s_1,s_2} A(\bk_n,s_1,s_2)\  \varphi^{\bks_n,s_1,s_2}(\bx)e^{-\Delta W_{\bks_n} t}
\end{eqnarray}
where $\Delta W_{\bks} = 2\sqrt{m_N^2+ \bk^2} - 2 m_N$. 
Using an identity $\Delta W_{\bks} = \bk^2/m_N - (\Delta W_{\bks})^2/(4m_N)$ and neglecting inelastic contributions, 
we arrive at the time-dependent Schr\"odinger-like equation
\begin{eqnarray}
\left\{-H_0 - \frac{\partial}{\partial t} +\frac{1}{4m_N}\frac{\partial^2}{\partial t^2}\right\}R(\bx,t) 
\label{eq:t-dep1}
&=& 
\int d^3 y U(\bx,\by) R(\by,t) \\
\label{eq:t-dep2}
&\simeq & V(\bx) R(\bx,t) +\cdots ,
\end{eqnarray}
which shows that the same $U(\bx,\by)$ in eq.~(\ref{eq:schroedinger}) can be obtained  from $R(\bx,t)$, where in the last line the derivative expansion is employed. 
An advantage of this extraction is that ground state saturation is not required for $R(\bx,t)$ to satisfy eq.~(\ref{eq:t-dep1}) or eq.~(\ref{eq:t-dep2}).
For this method to work, $t$ has to be large enough such that elastic contributions dominate $R(\bx,t)$. 

\subsection{Nucleon-Nucleon potentials}
\begin{figure}[h]
\begin{center}
  \includegraphics[height=0.48\textwidth,angle=-90]{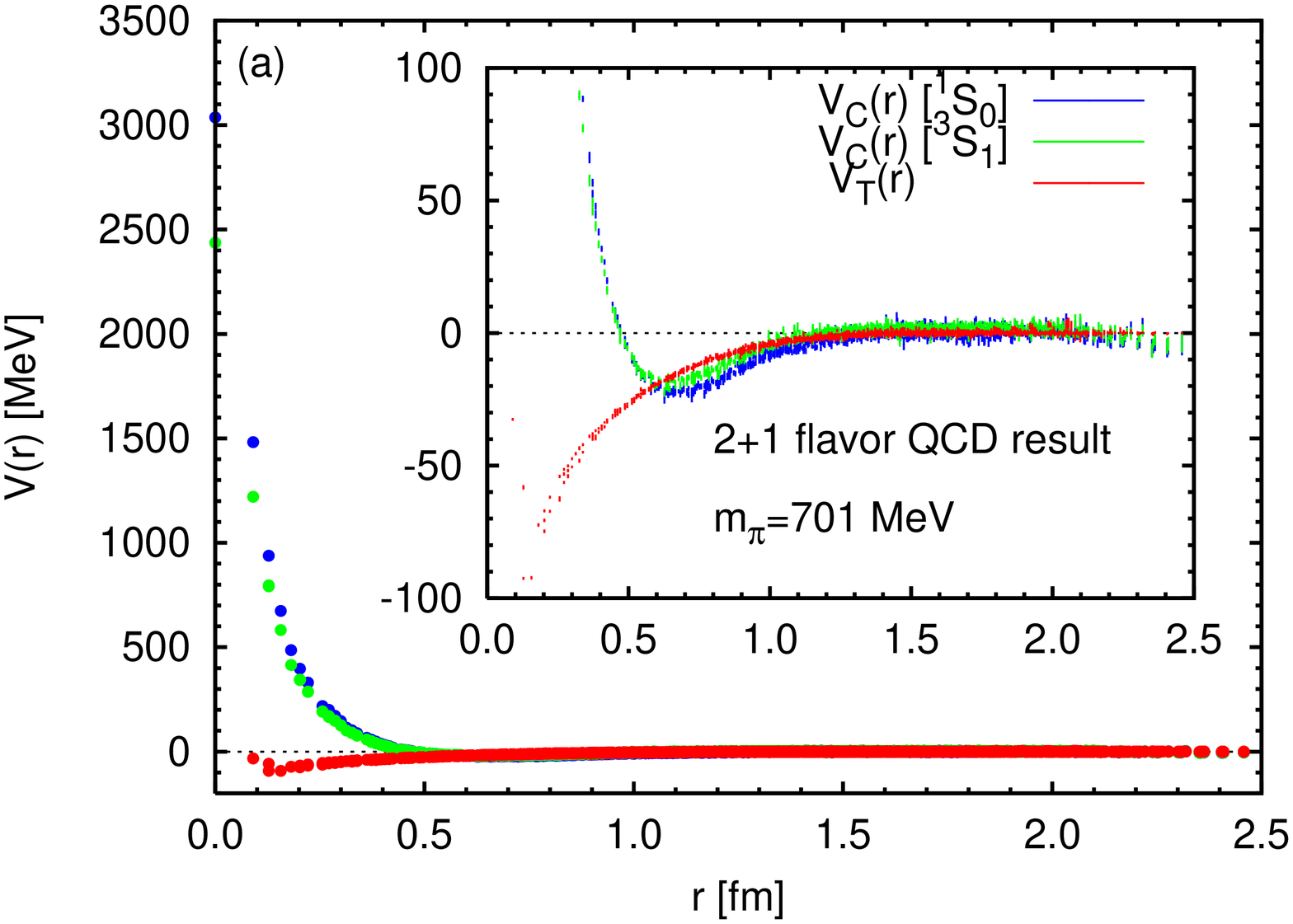}
  \includegraphics[height=0.48\textwidth,angle=-90]{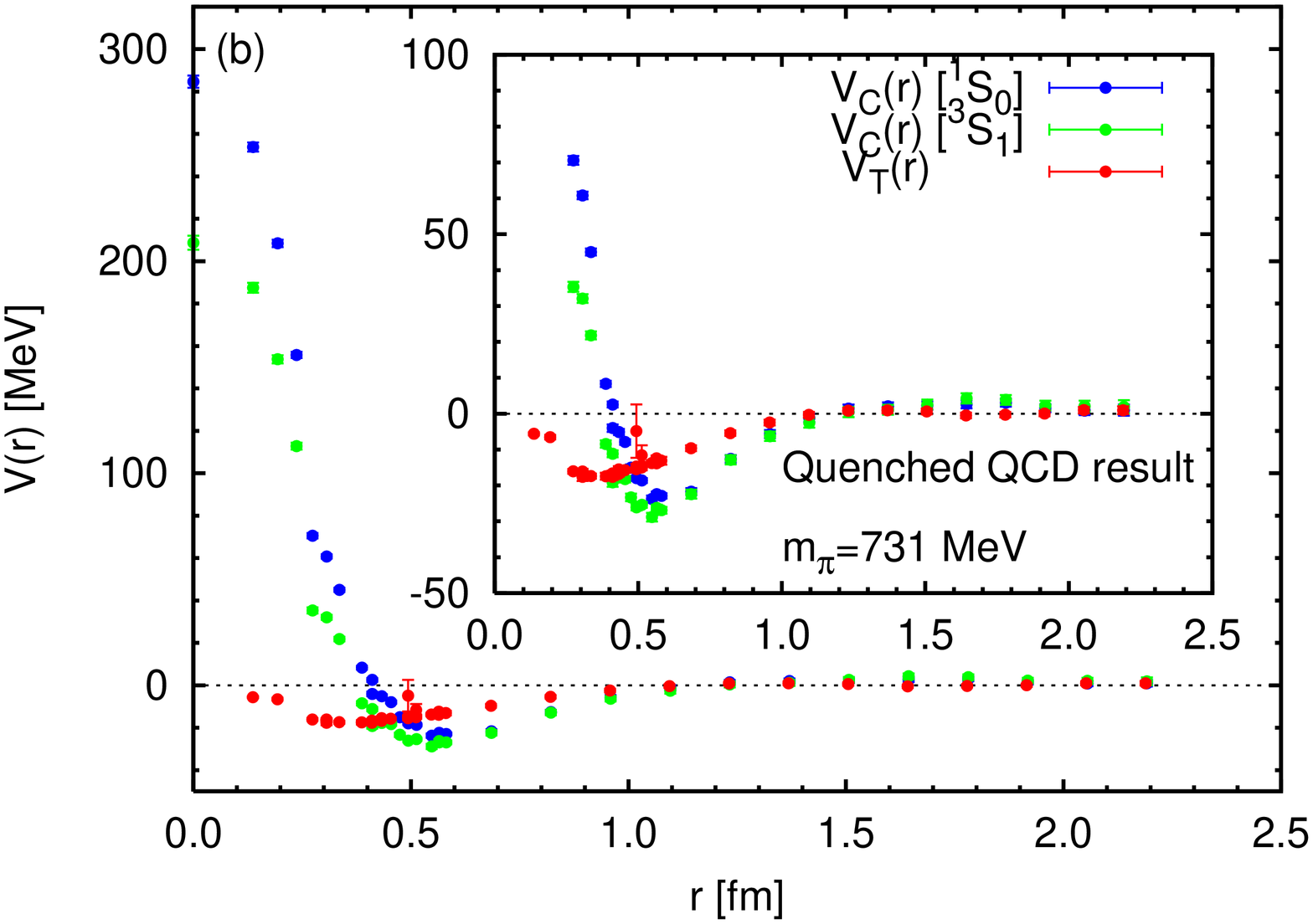}
\end{center}
\caption{ Central and tensor potentials in 2+1 flavor QCD at $m_{\pi} = 701$ MeV by the improved method (left) and 
 in quenched QCD at $m_{\pi} = 731$ MeV by the conventional method(right).}
\label{fig:NNpotential}
\end{figure}
We first show the parity-even NN potentials at the leading order in the derivative expansion in Fig.~\ref{fig:NNpotential},
where the central potential for the spin-singlet sector $V^{I=1}_c(r) \equiv V_0^{I=1}(r)-3V^{I=1}_\sigma (r)$, 
 the central potential for the spin-triplet sector $V^{I=0}_c(r) \equiv V^{I=0}_0(r)+V^{I=0}_\sigma (r)$ and
 the tensor potential for the spin-triplet sector $V_T^{I=0}(r)$ are plotted as a function of $r$ by blue, green and red symbols, respectively\cite{Ishii:2010th}. The results in the left figure have been obtained using PACS-CS gauge configurations  in the 2+1 flavor QCD at $m_\pi \simeq 701$ MeV and the lattice spacing $a\simeq 0.091$ fm on $32^3\times 64$ lattice\cite{Aoki:2008sm}, while those in the right figure have been obtained in quenched QCD at $m_\pi \simeq 731$ MeV and $a\simeq 0.137$ fm on a $32^4$ lattice.

The central potentials for both spin-singlet and spin-triplet sectors have repulsive cores, which confirm eq.~(\ref{V2N}) with positive $R$, 
while the tensor potential is less singular at short distance, in accordance with the fact that short distance contributions do not exist at the leading order of PT\cite{Aoki:2010kx}. Note however that the central potentials are finite at the origin in contrast to the divergent behavior predicted by the OPE analysis, probably due to the artifact of non-zero lattice spacing.
It is therefore important to show the appearance of the expected divergent behavior in future numerical simulations, by taking the continuum limit.   
We also observe that the repulsive cores and the tensor potential become significantly enhanced in the 2+1 flavor QCD. 

\subsection{Baryon interactions in the flavor SU(3) limit}
\begin{figure}[hbt]
\includegraphics[width=0.33\textwidth]{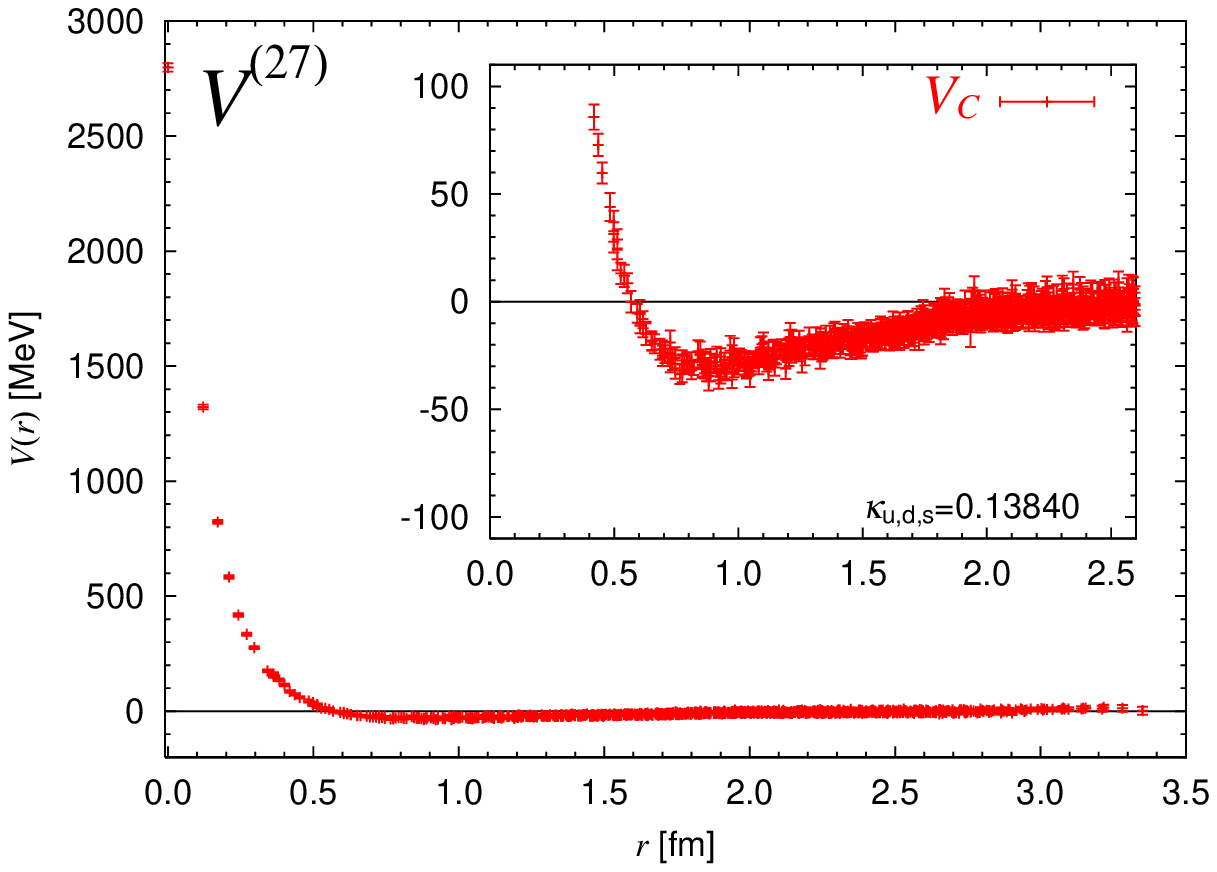} \hfill
\includegraphics[width=0.33\textwidth]{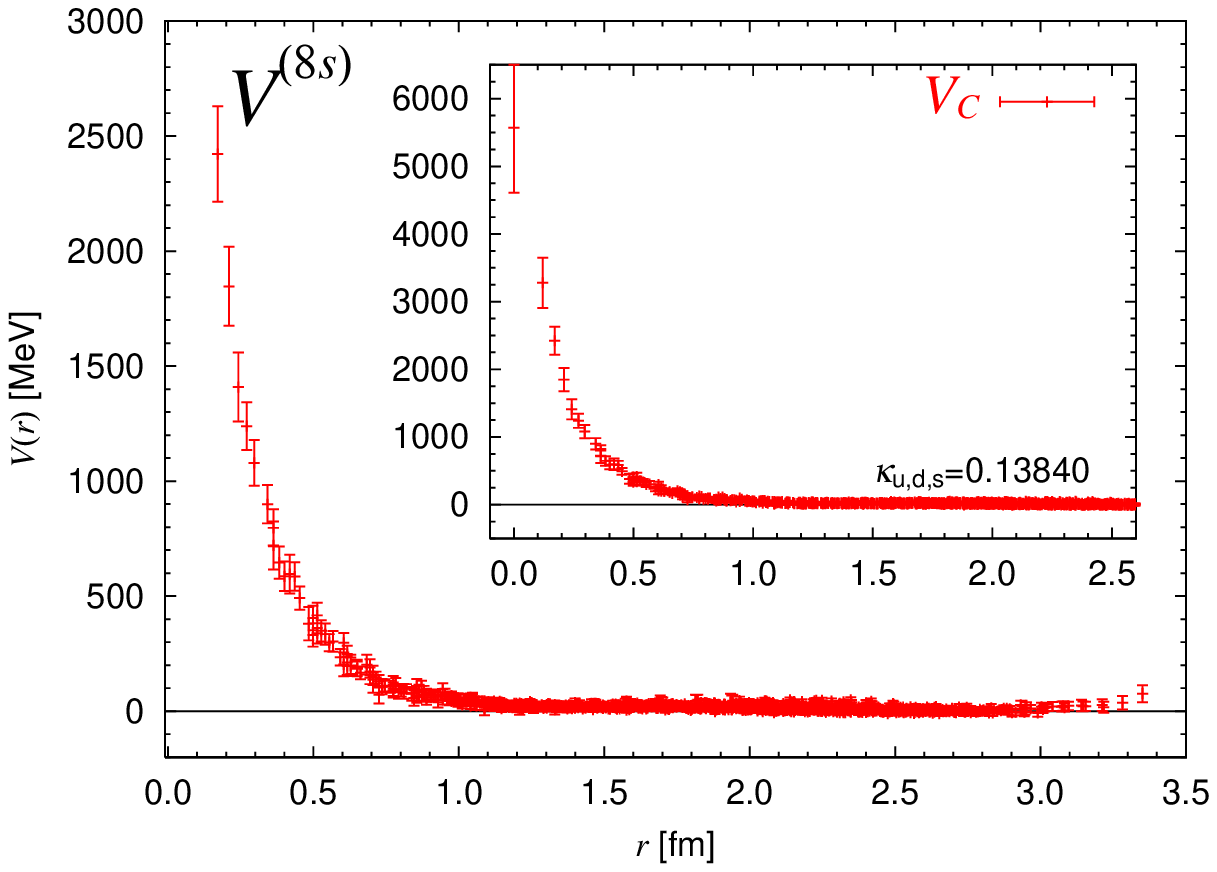} \hfill
\includegraphics[width=0.32\textwidth]{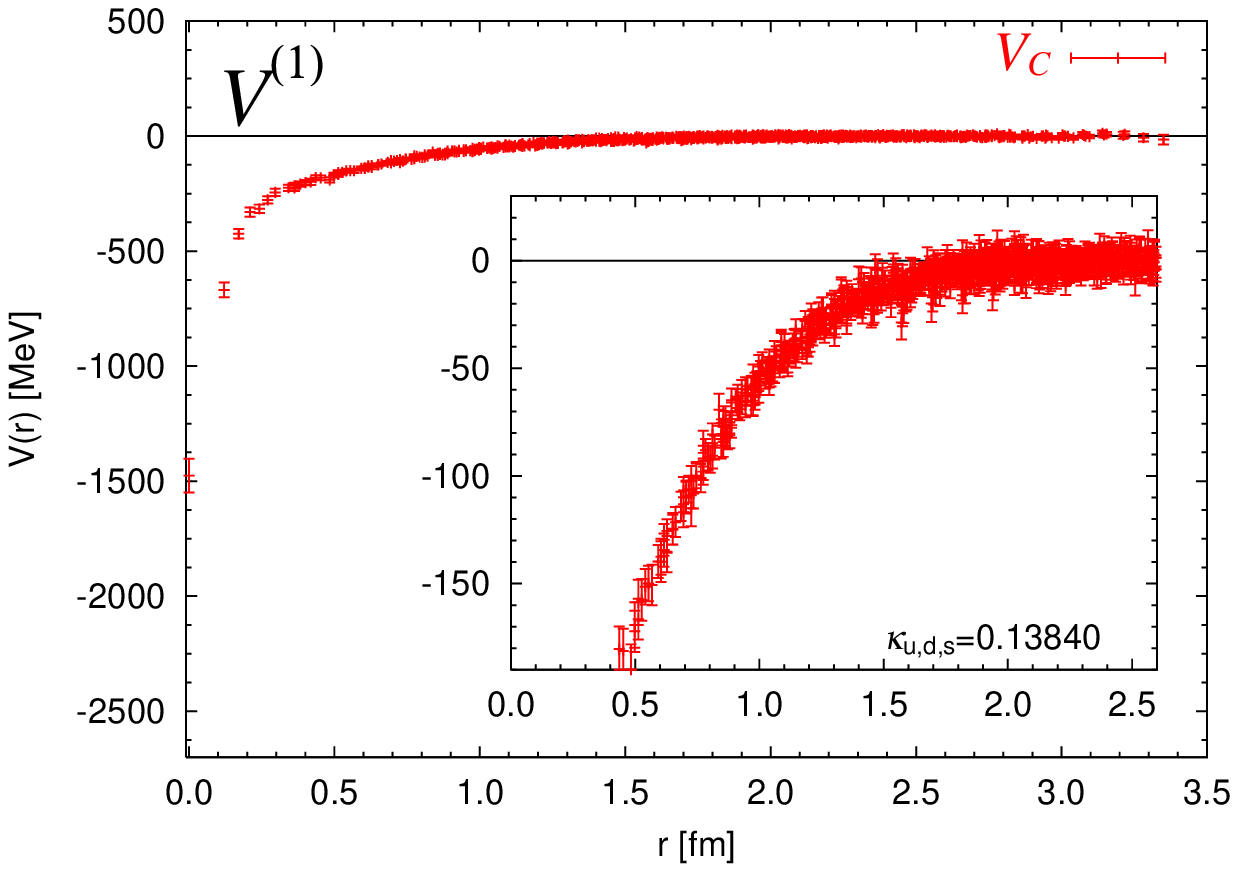} 
\smallskip\newline
\includegraphics[width=0.33\textwidth]{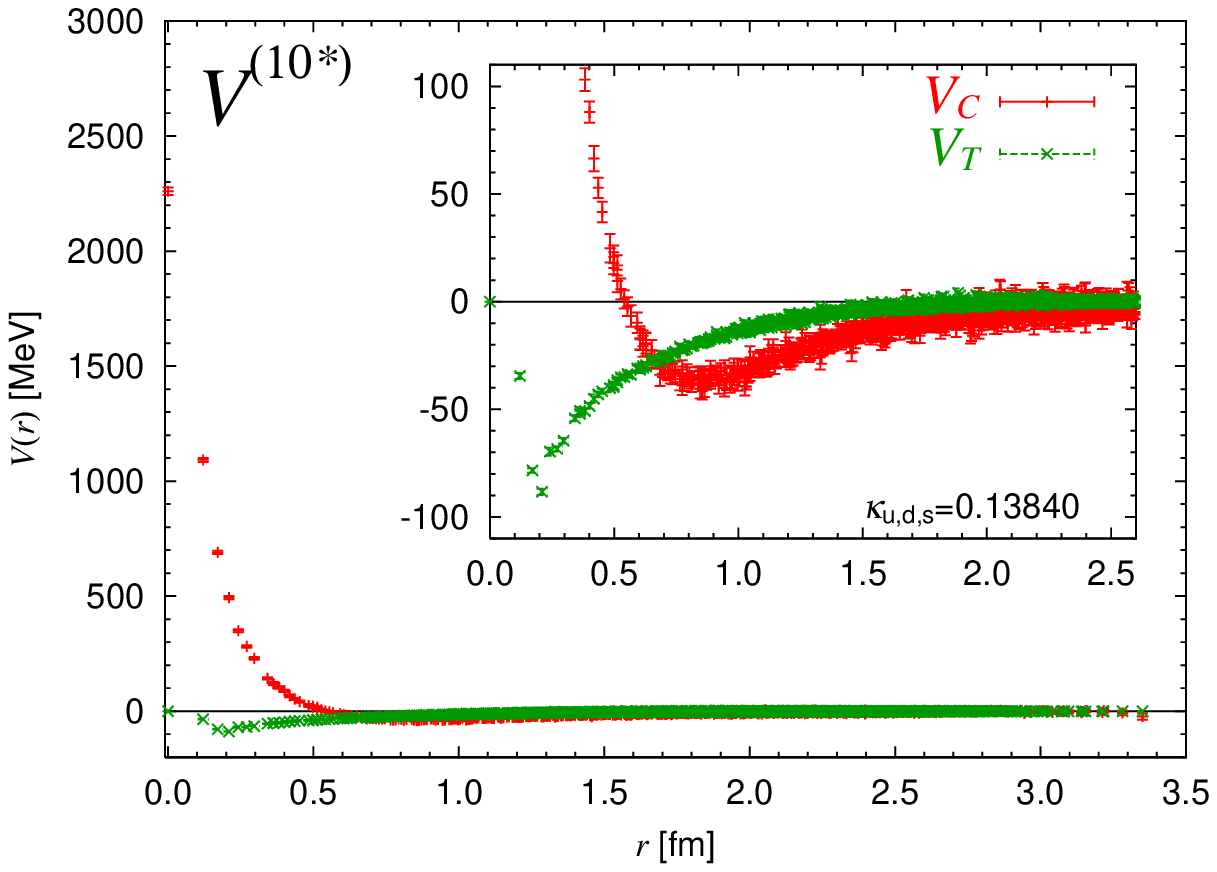}\hfill
\includegraphics[width=0.33\textwidth]{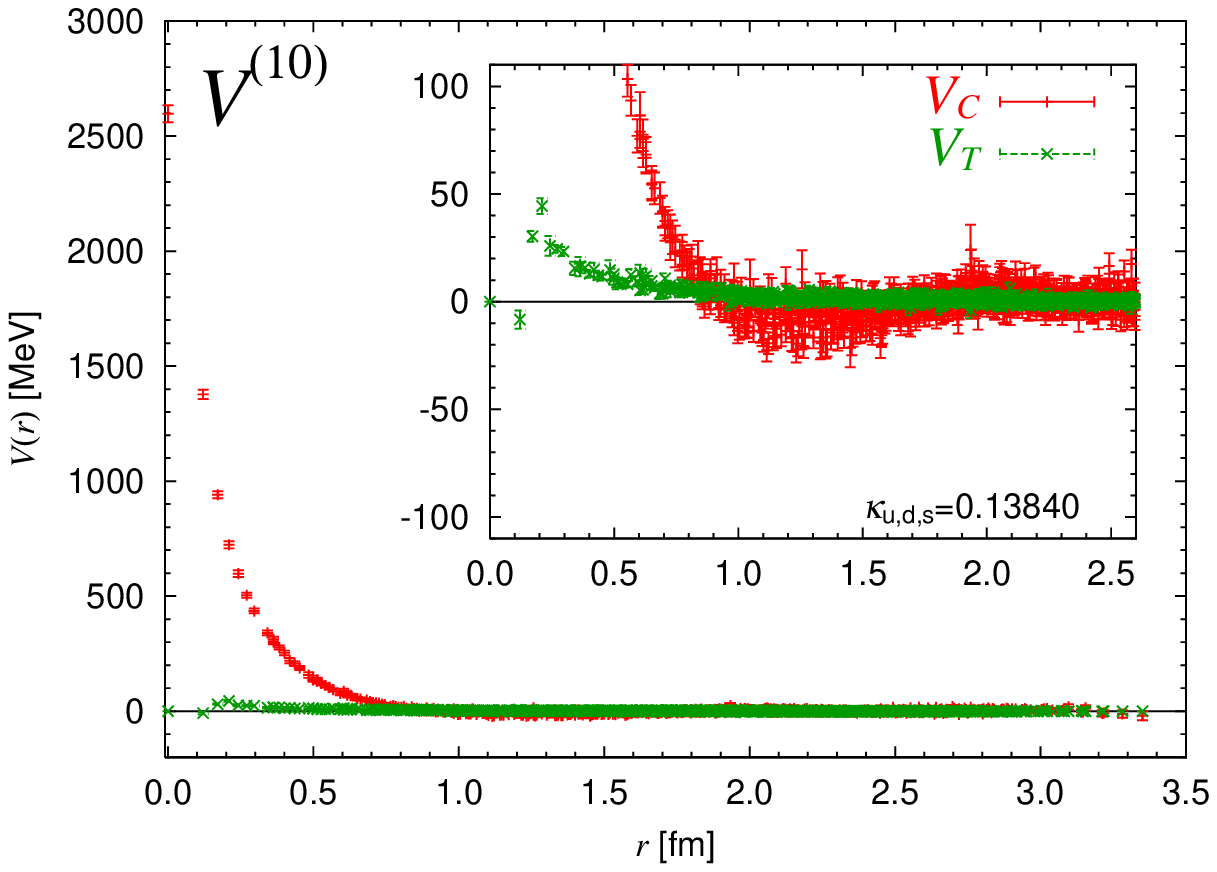}\hfill
\includegraphics[width=0.33\textwidth]{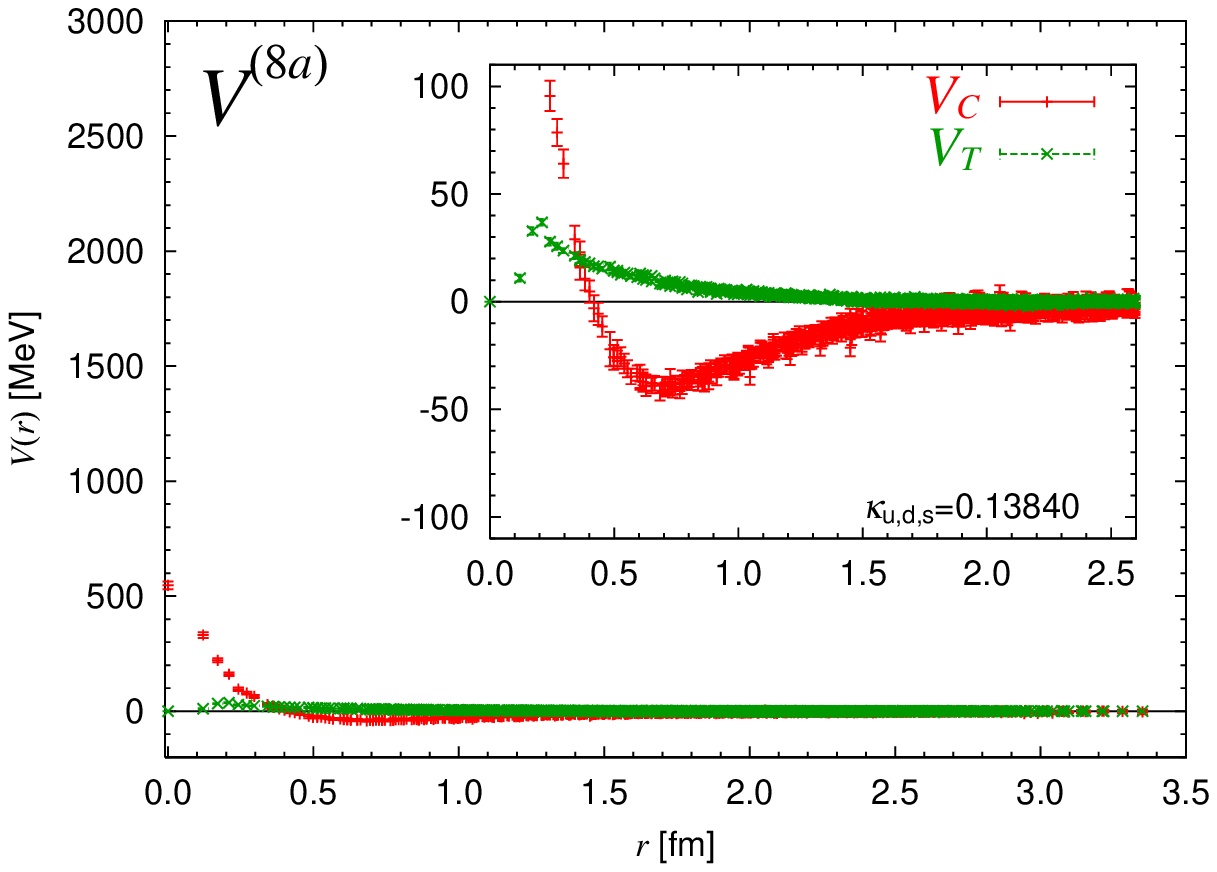}
\smallskip\newline
\caption{Potentials of baryon-baryon S-wave interaction in the flavor $SU(3)$ limit,
labeled by the flavor irreducible representation. These are obtained at  the pseudo-scalar meson mass of 469 MeV by the improved method.
}
\label{fig:potSU3}
\end{figure}
The figures in Fig.~\ref{fig:potSU3} show the leading order BB potential from full QCD simulations in the flavor SU(3) limit at the pseudo-scalar meson mass $M_{\rm PS} = 469(1)$MeV and $a=0.121(2)$fm\cite{Inoue:2011ai},
where the central potentials (red) in the spin-singlet sector are given for the ${27}$, ${8_s}$ and ${1}$ irreducible representations of SU(3), from left to right in the 1st row,  while the central and tensor potentials in
the spin-triplet sector are given for $\overline{10}$, ${10}$ and ${8_a}$ in the 2nd row.

The OPE analysis in the previous section predicts that, while the central potentials in the ${27}$, $\overline{10}$ and ${10}$ representations have a repulsive core, attractive interactions dominate at short distances
for the central potentials in the ${1}$, ${8_s}$ and ${8_a}$ representations.
Indeed, the central potential in the singlet representation, $V^{(1)}$, 
shows attraction at short distance in Fig.~\ref{fig:potSU3}, which leads to the bound H-dibaryon in this channel\cite{Inoue:2010es}.
The OPE analysis correctly predicts the attractive interaction at short distance in the singlet representation.
On the other hand, while the repulsion at short distance becomes weaker in the ${8}_a$ representation,
it becomes strongest among all 6 representations in the ${8}_s$ representation. The OPE analysis does not reproduce these repulsions.  

This puzzle may be resolved by observing \cite{Aoki:2010uz}
that there is no 6-quark operator in the $8_s$ channel in the
nonrelativistic limit and therefore
the corresponding nonperturbative matrix elements must be small.
Although the channel remains attractive at extremely short distances,
in the range between $0.1$ fm and $0.5$ fm relevant for lattice simulations
subleading operators, which are repulsive, dominate.
Note that a valence quark model with gluon exchange\cite{Oka:1983ku,Oka:1986fr} can reproduce the correct pattern of the short distance behavior of potentials observed in lattice QCD simulations including the attraction in the singlet channel.

\subsection{Three-nucleon force}
\begin{figure}[hbt]
\begin{center}
\includegraphics[width=0.49\textwidth]{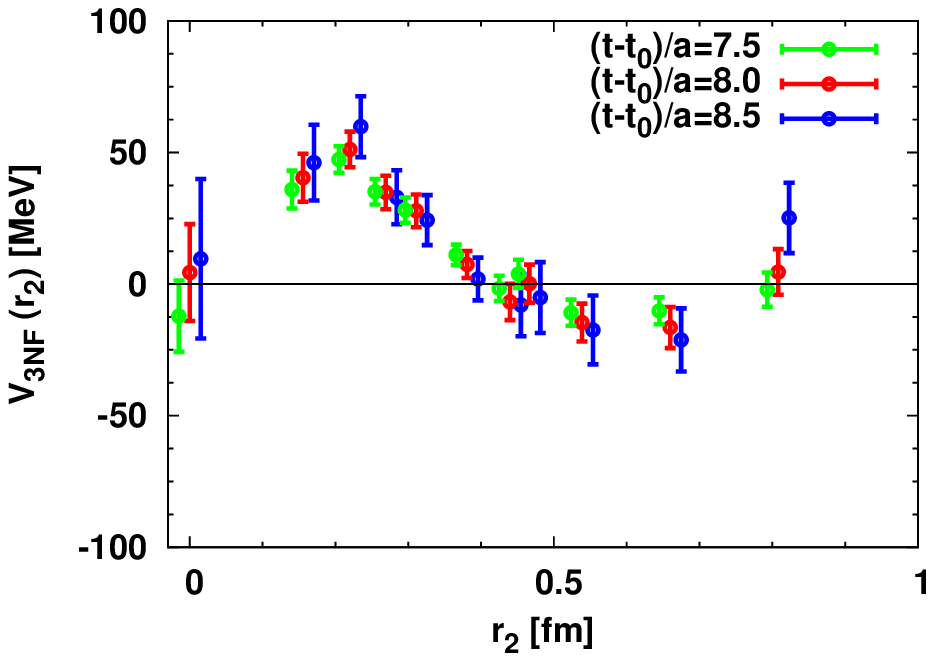}
\includegraphics[width=0.49\textwidth]{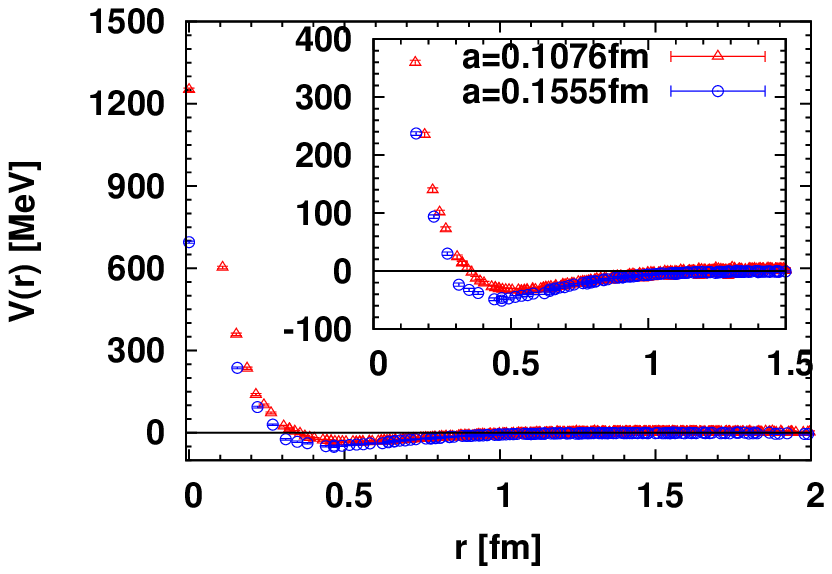}
\caption{
\label{fig:TNR}
(Left) The 3NF $V_{3N}(\br,\brho)$ at $\brho = 0$ as a function of $r_2=\vert\br\vert/2$, obtained by the improved method  at $(t-t_0)/a =7.5$(green), $8$(red) and $8.5$(blue) at $m_\pi  \simeq 1.13$ GeV and $a\simeq 0.156$ fm.
(Right)  The spin-singlet central potential in the 2-flavor QCD with $m_\pi \simeq 1.13$ GeV as a function of $r$ at $a\simeq 0.108$ fm (red) and 0.156 fm (blue). 
}
\end{center}
\end{figure}
The three-nucleon force (3NF) has been calculated in the linear setup with $\brho = 0$, where three nucleons 
are aligned linearly with equal spacing $r_2 = \vert \br \vert/2$, in 2-flavor QCD
at $a\simeq 0.156$ fm and $m_\pi \simeq 1.13$ GeV on a $16^3\times 32$ lattice\cite{Doi:2011gq}.
In the left of Fig.~\ref{fig:TNR}, we give the 3NF $V_{3N}(\br,\brho)$ at $\brho = 0$ as a function of $r_2$,
which are obtained by the improved method in eqs.~(\ref{eq:t-dep1}, \ref{eq:t-dep2}) for the 3NF\cite{Doi:2012am,Doi:2012ab} where $R(\bx,t)$ for two nucleons is replaced with $R(\br,\brho,t)$ for three nucleons.  
The results from the different sink times are consistent with each other within statistical errors. 
We observe an indication of repulsive 3NF at  short distance, which is consistent with the prediction by the OPE analysis that 3NF is universally repulsive.
As in the case of the 2NF, the potential is finite at the origin due to the non-zero lattice spacing in lattice QCD simulations. It is therefore important to further investigate the short distance behavior as a function of the lattice spacing $a$.
 
\section{Conclusions and discussions}
\label{sec:conclusion}
In this article, we have reviewed the recent activities to determine the short distance behavior of potentials among baryons, which are defined through the NBS wave function in the HAL QCD method.
We can determine the short distance behaviors of the NBS wave functions and thus the corresponding potentials,
employing the OPE (operator product expansion) and the RG analysis analytically in perturbation theory of QCD, thanks to its asymptotic freedom.  

The most notable predictions from the OPE analysis are 
(1) an existence of the repulsive cores for the two nucleon potentials  with the use of a simple effective model for the ratio of the matrix elements, 
(2) an appearance of the attractive core, instead of the repulsive one, for the flavor singlet channel in the 3-flavor QCD, and (3) an existence of  universal repulsion at short distance for the 3NF without relying on any non-perturbative inputs. These 3 results more or less agree with the behaviors of the corresponding potentials obtained in lattice QCD simulations, except the lattice potentials are always finite at the origin due to the lattice artifacts.
There are, however, some disagreements between the OPE analysis and lattice QCD results, which are probably caused by the effects of the neglected subleading terms in the OPE analysis at not asymptotically small distances.

The above success of the OPE analysis suggests that the repulsion among baryons at short distance may be explained by the combination of the Pauli suppression principle among quarks and the one-gluon exchange between quarks\footnote{The valence quark model with gluon exchange mentioned before relies on these properties\cite{Oka:1983ku,Oka:1986fr}.  }. Indeed in the OPE analysis, the tendency of repulsions become stronger for more numbers of quarks, while it becomes weaker for a larger number of valence flavors with the total number of quarks fixed.
Since the OPE analysis also predicts  forms of the repulsive/attractive cores as a function of distances, it is very important and interesting in the future to confirm them in lattice QCD simulations by taking the continuum limit. 
The right of Fig.~\ref{fig:TNR} shows our very preliminary result for a comparison of the spin-singlet central potential between two lattice spacings, $a\simeq $ 0.108 fm(red)  and 0.156 fm(blue).
We indeed observe an expected behavior that the repulsive core  increases as we decrease the lattice spacing.  More data and analysis in this direction will be required to determine the $r$ dependence of the  
repulsive cores.

\section*{Acknowledgements}
We would like to thank all members of HAL QCD collaboration for researches discussed in this report.
S. A. is supported in part by Grant-in-Aid for Scientific Research on 
Innovative Areas (No. 2004: 20105001,20105003) and by 
SPIRE (Strategic Program for Innovative Research).
T.D. is supported in part by 
MEXT Grant-in-Aid for Young Scientists (B) (24740146).
T.I. is supported in part by Grant-in-Aid for Scientific Research(C) 23540321.
This work  was also supported in part by the Hungarian National 
Science Fund OTKA (under K83267). 
S. A. and J. B. would like to thank the Max-Planck-Institut f\"ur
Physik for its kind hospitality during their stay for a part of this research.


\end{document}